\documentclass[a4paper,12pt, epsfig]{article}
\def\pr{0}

\usepackage{epsfig}
\usepackage{epstopdf}
\usepackage{graphicx}
\usepackage{ifthen}

\usepackage{amsmath}
\usepackage[psamsfonts]{amssymb}
\usepackage{euscript}

\usepackage{latexsym}

\newskip\humongous \humongous=0pt plus 1000pt minus 1000pt

\newif\ifdtup

\def\se{\hspace{-.07cm}=\hspace{-.07cm}}
\def\hsp{,\hspace{.7cm}}

\ifthenelse{\equal{\pr}{1}}{

\def\co{{\mathcal{O}}}

\renewcommand{\(}{\begin{equation}}
\renewcommand{\)}{end{equation} \vspace{-.05in}\linebreak}

\newcounter{saveeqn}
\newcounter{savealpheqn}

\newcommand{\alpheqn}{\setcounter{saveeqn}{\value{equation}}%
  \stepcounter{saveeqn}\setcounter{equation}{0}%
  \renewcommand{\theequation}{\mbox{\arabic{section}.\arabic{saveeqn}
\alph{equation}}}
  \renewcommand{\)}{\end{equation}}}
\def\part#1{\frac{\partial}{\partial{#1}}}%
\def\group#1{\refstepcounter{equation}\setcounter{saveeqn}
 {\value{equation}}%
  \label{#1}\setcounter{equation}{0}%
\renewcommand{\theequation}{\mbox{\arabic{section}.\arabic{saveeqn}
\alph{equation}}}
  \renewcommand{\)}{\end{equation}}}
\newcommand{\reseteqn}{\setcounter{equation}{\value{saveeqn}}%
  \renewcommand{\theequation}{\arabic{section}.\arabic{equation}}%
  \renewcommand{\)}{\end{equation}}}

\newcommand{\aalpheqn}{\setcounter{saveeqn}{\value{equation}}%
  \stepcounter{saveeqn}\setcounter{equation}{0}%
  \renewcommand{\theequation}{\mbox{
        \Alph{subsection}.\arabic{saveeqn}\alph{equation}}}
   \renewcommand{\)}{\end{equation}}}
\newcommand{\areseteqn}{\setcounter{equation}{\value{saveeqn}}%
  \renewcommand{\theequation}{\Alph{subsection}.\arabic{equation}}%
  \renewcommand{\)}{\end{equation}}}

\renewcommand{\thefootnote}{\alph{footnote}}
\renewcommand{\(}{\begin{equation}}
\renewcommand{\)}{\end{equation}}
\newcommand{\ba}{\begin{eqnarray}}
\newcommand{\ea}{\end{eqnarray}}

\newcommand{\bp}{\mathop{\vtop{\ialign{##\crcr
   $\hfil\displaystyle{}\hfil$\crcr\noalign{\kern-13pt\nointerlineskip}
   \BIG{(}\hskip0pt\crcr\noalign{\kern3pt}}}}}
\newcommand{\cbp}{\mathop{\vtop{\ialign{##\crcr
   $\hfil\displaystyle{}\hfil$\crcr\noalign{\kern-13pt\nointerlineskip}
   \BIG{)}\hskip0pt\crcr\noalign{\kern3pt}}}}}
\newcommand{\pa}{\mathop{\vtop{\ialign{##\crcr
    
$\hfil\displaystyle{\oplus}\hfil$\crcr\noalign{\kern+1pt\nointerlineskip 
}
   \hspace{.08in}$^{\alpha=0}$\hskip6pt\crcr\noalign{\kern3pt}}}}}
\renewcommand{\hsp}{,\hspace{.3in}}
\newcommand{\p}{^\prime}

\numberwithin{equation}{section}
\renewcommand{\theequation}{\mbox{\arabic{equation}}}

\def\df{\mathcal{D}_f}
\def\pin#1{\int \frac{d#1}{2\pi}}

\catcode`\@=11
\def\vereq#1#2{\lower3pt\vbox{\baselineskip1.5pt \lineskip1.5pt
\ialign{$\m@th#1\hfill##\hfil$\crcr#2\crcr\sim\crcr}}}
\catcode`\@=12

\renewcommand{\(}{\begin{equation}}
\renewcommand{\)}{\end{equation}}

\newcommand{\beas}{\begin{eqnarray*}}
\newcommand{\eeas}{\end{eqnarray*}}

\newcommand{\bquo}{\begin{quote}}
\newcommand{\enqu}{\end{quote}}

}{}

\newcommand{\beq}{\begin{equation}}
\newcommand{\eeq}{\end{equation}}
\newcommand{\bea}{\begin{eqnarray}}
\newcommand{\eea}{\end{eqnarray}}

\newskip\humongous \humongous=0pt plus 1000pt minus 1000pt

\newif\ifdtup

\newcommand{\p}{^\prime}

\def\co{{\mathcal{O}}}

\def\df{\mathcal{D}_f}
\def\pin#1{\int \frac{d#1}{2\pi}}

\def\noprl#1{\ifthenelse{\equal{\pr}{1}}{}{#1} }

\begin{document}
% ======================================================================== 
\def\thefootnote{\fnsymbol{footnote}}
\def\thetitle{Well-Defined Quantum Soliton Masses Without Supersymmetry}
\def\autone{Jarah Evslin}
\def\auttwo{Baiyang Zhang}
\def\affa{Institute of Modern Physics, NanChangLu 509, Lanzhou 730000, China}
\def\affb{University of the Chinese Academy of Sciences, YuQuanLu 19A, Beijing 100049, China}

\ifthenelse{\equal{\pr}{1}}{
\title{\thetitle}
\author{\autone}
%\author{\auttwo}
\affiliation {\affa}
\affiliation {\affb}
}{
\begin{center}
{\large {\bf \thetitle}}

\bigskip

\bigskip

{\large \noindent  \autone \footnote{jarah@impcas.ac.cn}}

\vskip.7cm

1) \affa\\
2) \affb\\

\end{center}
}

\begin{abstract}
\noindent

\noindent
22 years ago, Rebhan and van Nieuwenhuizen showed that loop corrections to the mass of a quantum soliton depend on a choice of matching condition for the regulators of the vacuum and one-soliton sector Hamiltonians.  In supersymmetric theories, regulators which preserve supersymmetry yield the correct quantum corrections, as these are dictated by supersymmetry.  However, in a general theory it is not known which matching condition yields the correct mass.  We use the leading term in the operator that creates the soliton to construct the regulated one-soliton sector Hamiltonian from that of the vacuum sector, providing the correct matching condition.  As an application, we derive a simple formula for the one-loop mass of a kink in a large class of 1+1 dimensional scalar field theories and also, at one loop, we diagonalize the Hamiltonian which describes the kink excitations.

%We present a simple formula for the one-loop correction to the mass of a kink in any 1+1 dimensional scalar field theory with a canonical kinetic term and a non-derivative scalar potential that satisfies a simple reflection symmetry condition.  At one loop we also diagonalize the kink sector Hamiltonian.  Previous such calculations have depended on an arbitrary matching of the regulators of the vacuum and one kink sector Hamiltonians.  Instead here, using the operator which creates the kink, once the vacuum sector is renormalized, the renormalization of the kink sector is automatic and well-defined.   For simplicity, the renormalization of the vacuum sector is achieved by normal ordering,  which renders it ultraviolet finite.

\end{abstract}

% \vfill
% 
% \end{titlepage}
\setcounter{footnote}{0}
\renewcommand{\thefootnote}{\arabic{footnote}}

% \pacs{??}

\ifthenelse{\equal{\pr}{1}}{
\maketitle
}{}

%\pr
%\ifthenelse{\equal{\pr}{1}}{yippy}

Quantum solitons play a critical role in many fields, from the vortices inside of superconductors and neutron star superfluids to the monopoles or center vortices that may be responsible for confinement in quantum chromodynamics to the branes whose identification sparked the second superstring revolution.  In the semiclassical regime, the mass of a quantum soliton can be calculated in a loop expansion.  This expansion requires both the vacuum sector Hamiltonian $H$ and the one soliton sector Hamiltonian $H_K$ to be simultaneously regularized and renormalized.  The regulators must then be taken to infinity.  However, in Ref.~\cite{problema} it was shown that the relationship between the regulators when they are taken to infinity affects the answer, with some prescriptions yielding the correct quantum mass and some yielding incorrect masses.   

In this letter we argue that the following prescription yields the correct matching condition.  Let $\df$ be the unitary operator which shifts the values of the fields from a vacuum value to the classical soliton solution.  Then the two regularized Hamiltonians are related by the similarity transformation
\beq
H_K=\df^{-1} H \df.
\eeq
This definition completely specifies the regularized $H_K$ given the regularized $H$.  Within the validity of the semiclassical expansion, the soliton masses can now be computed perturbatively using $H_K$.  We perform this exercise for a broad class of kink solutions, recovering a general expression for the mass at one loop which agrees with that of Ref.~\cite{dhn2} for the $\phi^4$ kink and \cite{rajaraman} for the Sine-Gordon soliton. 

The mass of a soliton in a quantum field theory is the difference between the lowest energies $E_K$ and $E_0$ of states in the one-soliton and vacuum sectors respectively.  In a quantum field theory, both of these numbers are generally infinite.  The infrared divergences that arise from constant density at infinite volume are easily treated, for example by adding a constant to the Hamiltonian density.  More troublesome are the ultraviolet divergences.    The standard approach to these \cite{dhn2,rajaraman,physrept04} is to separately regularize both the soliton and vacuum sectors with regulators parametrized by $r_K$ and $r_0$ respectively, and then add counterterms to each which are fixed using a renormalization condition.  The soliton mass is thus
\beq
M_s=\stackrel{lim}{{}_{r_K,r_0\rightarrow \infty}}E_K^{r_K}-E_0^{r_0}+c.t. \label{loro}
\eeq 
where regulator-dependent counterterms {\it{c.t.}} have been included and subtracted in the soliton and vacuum sectors respectively.  

Eq.~(\ref{loro}) is quite ill-defined.  While the counter-terms can and are chosen to eliminate ultraviolet divergences, the remaining finite contributions depend on the relation between $r_0$ and $r_K$ when they are both taken to infinity.   The simplest regularization schemes correspond to removing certain ultraviolet modes, and so this relation between $r_0$ and $r_K$ corresponds to an identification of cutoffs in the vacuum and soliton sectors.  There are many choices of identification.  For example, in the case of the $\phi^4$ kink in 1+1 dimensions, Ref.~\cite{dhn2} compactified their theory so that their spectra were discrete and identified modes by pairing them in order of increasing energy.   The motivation for our work comes from Ref.~\cite{problema}, which showed that matching based on mode counting yields a different finite contribution to the mass than a matching based on a fixed UV energy cutoff.  

The fact that the expression (\ref{loro}) depends on the matching prescription already at one loop led researchers to wonder which matching prescription, if any, is correct.  More prescriptions were invented \cite{n2,glocal,wtesi} and the results were checked against expectations from integrability and supersymmetry.   In Ref.~\cite{n2} the authors proposed a general definition of the kink mass in the special case of a symmetric potential.  It requires a compactification of size $L$ as well as a massless limit, and it is suggested that a double scaling limit of $L$ and the mass exists in which finite mass effects do not affect the results.  In Ref.~\cite{schon} a distant anti-soliton was added, and the author argues that as a result the boundary conditions with the pair are the same as the vacuum sector and so the ambiguity should vanish.  Once the dust settled, it was concluded \cite{physrept04} that some matching prescriptions provide the expected results from integrability and supersymmetry at one-loop, while others are ``bad" \cite{bad}.   In the supersymmetric case, it was possible to obtain the correct mass corrections by using a regulator that preserves the supersymmetry \cite{susy}.  However it remained unclear whether the successful prescriptions continue to produce the correct answer in the absence of integrability or supersymmetry or, more generally, beyond one loop.

In this letter, we will use an operator which takes the vacuum sector to the one-kink sector to derive the correct matching prescription, which can then be used reliably to calculate the mass of a kink to any order in perturbation theory.  As an application, we will calculate the mass of a kink in a fairly general class of scalar theories at one-loop.   As an illustration, the case of the kink in the $\phi^4$ theory is worked out in gory detail in the companion paper~\cite{mekink}.

We will consider a (1+1)-dimensional theory of a scalar field $\phi(x)$ with Hamiltonian 
\bea
H&=&H_0+H_I\hsp H_I=\int dx:\left[\frac{M^2}{g^2}V(g\phi(x))
\right]:\nonumber\\
H_0&=&\int dx:\left[\frac{1}{2}\pi(x)\pi(x)+\frac{1}{2}\partial_x\phi(x)\partial_x\phi(x)\right]:
\eea
where $\pi(x)$ is the conjugate momentum to $\phi(x)$ and the normal-ordering will be defined momentarily.  The constant $g$ has units of action${}^{-1/2}$ and so always appears in the dimensionless combination $g\hbar^{1/2}$.  This implies that our loop expansion is an expansion in $g^2\hbar.$   We will set $\hbar=1$.  

Consider two adjacent minima $y_1$ and $y_2>y_1$ of the potential $V$ such that
\beq
V(y_1+y)=V(y_2-y)\rm{\ \ if\ \ }y\in[0,y_2-y_1]. \label{sym}
\eeq
For convenience, the potential may always be shifted so that $V(y_1)=0$.  

Define the shifted field
\beq
\tilde{\phi}(x)=\phi(x)-\frac{y_1}{g}.
\eeq
Note that the shifted field is canonically conjugate to the original $\pi(x)$.  Expanding the Hamiltonian to second order in $\tilde{\phi}$, one sees that it is a scalar field of mass squared
\beq
m^2=M^2\partial^2_yV(y)|_{y=y_1}.
\eeq
In the Schrodinger picture, the shifted field and its conjugate ${\pi}(x)$ can be decomposed as usual into oscillator modes $a^\dag_p$ and $a_p$ which satisfy a Heisenberg algebra
\bea
\tilde{\phi}(x)&=&\pin{p}\frac{1}{\sqrt{2\omega_p}}\left(a^\dag_p+a_{-p}\right)e^{-ipx}
\nonumber\\
{\pi}(x)&=&i\pin{p}\frac{\sqrt{\omega_p}}{\sqrt{2}}\left(a^\dag_p-a_{-p}\right)e^{-ipx} \label{osc}
\eea
where
\beq
\omega_p=\sqrt{m^2+p^2}.
\eeq
The normal ordering is defined with respect to these modes.  Recall that in (1+1)-dimensional scalar field theories with such nonderivative potentials, normal ordering is sufficient to render the theory finite in the ultraviolet.  In a sense, the normal ordering corresponds to adding a certain choice of counterterm after which regularization is no longer required.  We claim that the calculation below would proceed similarly with a more general counterterm and regulator, and so can also be applied to more interesting field theories.  However the proof of this claim will be left to future work.

The classical field theory admits a stationary kink solution $\tilde{\phi}(x,t)=f(x)$ which solves
\beq
f^{\prime\prime}(x)=\frac{M^2}{g}V\p(gf(x)+y_1)
\eeq
with boundary conditions $f(-\infty)=0$\ and $f(+\infty)=y_2-y_1$.   If $f_1(x)$ is the solution that would be obtained at $g=1$, then $f(x)=f_1(x)/g$.  In this sense $f$ is of order $O(1/g)$.

Let us define the displacement operator
\beq
\df={\rm{exp}}\left(-i\int dx f(x){\pi}(x)\right) \label{df}
\eeq
which satisfies the relation \cite{mekink}
\beq
:F\left({\pi}(x),\tilde{\phi}(x)\right):\df=\df:F\left({\pi}(x),\tilde{\phi}(x)+f(x)\right): \label{fident}
\eeq
and is unitary.  It defines the new Hamiltonian $H_K$ via the definition
\beq
H\df=\df H_K. \label{hk}
\eeq

The key observation behind this letter is that (\ref{hk}) can be used to uniquely define the operator $H_K$ however $H$ is regulated and whatever counterterms have been added to $H$.  This provides the matching prescription advertised above.  In the case at hand
\beq
H_K=H_0+\int dx:\left[\frac{M^2}{g^2}V(g\tilde{\phi}(x)+gf(x)+y_1)
\right]:.
\eeq
If $V$ is analytic at $y_1$, we may expand $V$ and so $H_K$ in a power series in $\tilde\phi$.  All terms of order $\tilde{\phi}^k$ are multiplied by $g^{k-2}$.  Thus the cubic terms and above may be treated in perturbation theory in $g$.   At order $O(g^0)$ we find the truncated Hamiltonian~$H_{K0}$
\bea
H_K&=&H_{K0}+E_{\rm{cl}}+O(g)\label{trunc}\\
E_{\rm{cl}}&=&\int dx \frac{f\p(x)^2}{2}+\frac{M^2}{g^2}V(f_1(x)+y_1)\nonumber\\
H_{K0}&=&H_0+\frac{M^2}{2}\int dx V^{\prime\prime}(f_1(x)+y_1):\tilde{\phi}(x)^2:.\nonumber
\eea
Here $E_{\rm{cl}}$ is the classical kink mass.  By dimensional analysis $E_{\rm{cl}}$ is of order $O(M/g^2)$.  Therefore the $g$-independent correction that one may expect from $H_{K0}$ is suppressed by $g^2\hbar$ with respect to $E_{\rm{cl}}$ and so is the one-loop correction.  The $O(g)$ corrections on the first line then only enter at two loops and beyond.  What is $H_{K0}$?

The energy of a state is its eigenvalue with respect to the Hamiltonian $H$.  Let us consider two eigenstates of $H$, the vacuum $|0\rangle$ with $\langle 0|\tilde{\phi}(x)|0\rangle=0$ and the ground state of the kink sector, hereafter called the kink state $|K\rangle$.  The mass $M_K$ of the kink is just the difference between their two eigenvalues
\beq
E_0|0\rangle=H|0\rangle,\ \  E_K|K\rangle=H|K\rangle,\ \  M_K=E_K-E_0. \label{eig}
\eeq
Note that the same operator $H$ appears in each expression, so there is only one operator to regularize and so no need to match regulators.

Following the logic of \cite{hepp}, $\df|0\rangle$ is in the one-kink sector because $\langle0|\df^{-1}\tilde{\phi}(x)\df|0\rangle=f(x)$.   However it is not the kink ground state $|K\rangle$, indeed it is not even an eigenstate of the Hamiltonian.   As argued in Ref.~\cite{taylor78}, the kink ground state will contain loop corrections to this naive formula.  We will encode these corrections in an operator $\mathcal{O}$ defined to be any operator which satisfies
\beq
|K\rangle=\df\co|0\rangle.
\eeq
Then, using Eq.~(\ref{eig})
\beq
E_K\co |0\rangle\se\df^{-1}E_K\df\co|0\rangle\se\df^{-1}H\df\co|0\rangle\se H_K\co|0\rangle. \label{neweq}
\eeq
That $E_K$ is an eigenvalue of $H_K$ is obvious, as $H_K$ and $H$ are related by a similarity transformation and so have the same eigenvalues.   The lowest eigenstate of $H_K$ is $\df^{-1}\co |0\rangle$ which has eigenvalue $E_0$.  

$\co$ can be taken to be equal to the identity plus perturbative loop corrections.  This means that one may truncate $H_K$ to any desired order in $g$ and find $\co$ and $E_K$ at the same order.  Inserting (\ref{trunc}) expanded to order $O(g^0)$, Eq.~(\ref{neweq}) becomes the eigenvalue problem
\beq
H_{K0}\co |0\rangle=(E_K-E_{\rm{cl}})\co |0\rangle
\eeq
which can be solved by diagonalizing $H_{K0}$.   The free Hamiltonian $H_{K0}$ describes the linearized theory of local perturbations to the kink and so its eigenstates are all in the one-kink sector.  Thus the lowest energy eigenstate of $H_{K0}$ is the kink ground state $\co|0\rangle$.   To find it, we will now diagonalize $H_{K0}$.

At leading order the classical equation of motion derived from $H_{K0}$ is
\beq
-\partial_t^2 F(x,t)+\partial_x^2 F(x,t) =M^2V^{\prime\prime}(gf(x)+y_1)F(x,t).
\eeq
Decomposing $F(x,t)=F(x)e^{-i\omega t}$ one obtains
\beq
\partial_x^2 F(x) =\left(-\omega^2+M^2V^{\prime\prime}(gf(x)+y_1)\right)F(x).
\eeq
We will be interested in such solutions which remain bounded as $x\rightarrow\pm\infty$.  Defining
\beq
V_I(\tilde{\phi})=V(\tilde\phi)-\frac{m^2}{2}\tilde{\phi}^2\hsp k^2=\omega^2-m^2
\eeq
this becomes
\beq
\partial_x^2 F(x) =\left(-k^2+M^2V_I^{\prime\prime}(gf(x)+y_1)\right)F(x) \label{feq}
\eeq
There will be continuum solutions and also $\beta+1$ bound state solutions, which we will index by $\mu\in[0,\beta]$ with $\mu=0$ corresponding to the Goldstone mode.  Due to the condition (\ref{sym}) the bound state solutions $F_{B,\mu}$ will be even or odd while the continuum solutions $F_k$ at each $k$ can be decomposed into an even and odd part.  We will organize all solutions $F(x)$, bound or unbound, such that the real part is even and the imaginary part is odd $F^*(x)=F(-x)$ and we will normalize them so that
\bea
\int dx F_k(x) F^*_l(x)dx&=&2\pi\delta(k-l)\nonumber\\
\int dx F_{B,\mu}(x) F^*_{B,\nu}(x)dx&=&\delta_{\mu\nu}.
\eea

Next we decompose $\tilde\phi$ and $\pi$ into contributions from the continuum and bound states
\[
\phi(x)\se\phi_C(x)+\sum_{\mu=0}^\beta \phi_{B,\mu}(x),\ \pi(x)\se\pi_C(x)+\sum_{\mu=0}^\beta\pi_{B,\mu}(x)
\]
which in turn are decomposed in terms of solutions
\bea
\phi_C(x)&=&\pin{k}\frac{1}{\sqrt{2\omega_k}}\left(b_k^\dag+b_{-k}\right)F_k(x) \label{bdec} \\
\phi_{B,i}(x)&=&\frac{1}{\sqrt{2\omega_{B,i}}}\left(b_{B,i}^\dag-P_i b_{B,i}\right)F_{B.i}(x)\nonumber\\
\phi_{B,0}(x)&=&\phi_0 F_{b,0}(x),\ \pi_{B,0}(x)=\pi_0 F_{B,0}(x)\nonumber\\
\pi_C(x)&=&i \pin{k}\sqrt{\frac{\omega_k}{2}}\left(b_k^\dag - b_{-k}\right)F_k(x)\nonumber\\
\pi_{B,i}(x)&=&i \sqrt{\frac{\omega_{B,i}}{2}}\left(b_{B,i}^\dag+P_i b_{B,i}\right)F_{B,i}(x)\nonumber
\eea
where the index $i\in[1,\beta]$ runs over all bound states except for the Goldstone mode $\mu=0$, which has frequency $\omega_{B,0}=0$ and the sign $P_i$ is the parity of the $i$th bound state.  

Combining (\ref{osc}) and (\ref{bdec}) one can find the Bogoliubov transformation relating the two sets of oscillators
\bea
a_p&=&a_{C,p}+\sum_{\mu=0}^\beta a_{B,\mu,p}\label{bog}\\
a_{C,p}&=&\pin{k}\frac{\hat F^*_k(p)}{2}\left(\frac{\omega_p-\omega_k}{\sqrt{\omega_p\omega_k}}b_k^\dag+\frac{\omega_p+\omega_k}{\sqrt{\omega_p\omega_k}}b_{k}\right)\nonumber\\
a_{B,i,p}&=&\frac{\hat F^*_{B,i}(p)}{2}\left(\frac{\omega_p-\omega_{B,i}}{\sqrt{\omega_p\omega_{B,i}}}b_{B,i}^\dag+P_i \frac{\omega_p+\omega_{B,i}}{\sqrt{\omega_p\omega_{B,i}}}b_{B,i}\right)\nonumber\\
a_{B,0,p}&=&\hat F^*_{B,0}(p)\left[ \sqrt{\frac{\omega_p}{2}}\phi_0+\frac{i}{\sqrt{2\omega_p}}\pi_0\right]\nonumber
\eea
where $\hat F$ is the Fourier transform of $F$.  This Fourier transform will generally contain a term proportional to $\delta(k-p)$ and, if the potential in $H_{K0}$ is not reflectionless, also $\delta(k+p)$.  These delta functions can be used to evaluate the $k$ integrals in their respective terms.  As $\omega_p=\omega_{-p}$, one sees from (\ref{bog}) that these terms do not lead to mixing between $a$ and $b^\dagger$.  They therefore preserve the normal ordering and so will not contribute to soliton masses.

Recall that $H_{K0}$ in Eq.~(\ref{trunc}) is normal ordered in terms of $a$ and $a^\dag$, so that all $a^\dag$ appear on the left.  Inserting (\ref{bog}) it may be written in terms of $b^\dag$ and $b$.  By direct computation one can see that there are no $b^\dagger b^\dagger$ or $bb$ terms.  However there will be terms of the form $b b^\dagger$. By inverting (\ref{bdec}) and using the canonical commutation relations of $\tilde\phi$ and $\pi$ one finds that $b$ satisfy the algebra
\[
[b_{k_1},b^\dag_{k_2}]\se 2\pi\delta(k_1-k_2),\ 
[b_{B,i},b^\dag_{B,j}]\se \delta_{ij},\  [\phi_0,\pi_0]\se i.
\]
Using these commutation relations and the defining equation (\ref{feq}) of $F$, $H_{KO}$ may be reordered to place all $b^\dagger$ on the left, yielding
\bea
H_{K0}&=&\pin{k}\omega_k b^\dag_k b_k+\sum_i \omega_{B,i} b^\dag_{B,i} b_{B,i}
+\frac{\pi_0^2}{2}+Q\nonumber\\
Q&=&-\frac{1}{4}\left[\pin{k}\pin{p}\frac{(\omega_p-\omega_k)^2}{\omega_p}\hat F_k^2(p)
\right.\nonumber\\&&\left.+\sum_{\mu=0}^\beta\pin{p}\frac{(\omega_p-\omega_{B,\mu})^2}{\omega_p}\hat F_{B,\mu}^2(p)
\right].
\eea
This is just a sum of harmonic oscillators.  

Therefore the lowest energy mode, $\co|0\rangle$, is the one annihilated by all $b$
\beq
b_k \co |0\rangle=b_{B,i}\co |0\rangle=\pi_0\co |0\rangle=0
\eeq
and its corresponding energy is $E_K=E_{\rm{cl}}+Q$.  The one-loop correction to the kink mass is just $Q$.  In Refs.~\cite{mekink} and \cite{hengyuan} we show that, in the case of the $\phi^4$ kink and Sine-Gordon soliton, this agrees with the one-loop result obtained using mode matching but not that obtained using an energy cutoff.  

This is our main result.  However by diagonalizing the Hamiltonian we have obtained the entire spectrum, at 1-loop.  It consists of harmonic oscillator spectra excited with various $b^\dagger$ and also rigid momenta $\pi_0$.  To calculate the higher loop corrections is now straightforward.   One need only consider the full Hamiltonian in Eq.~(\ref{trunc}) and use ordinary perturbation theory in $g^2$, in terms of the $b$ and $b^\dagger$ oscillators.  At 2 loops there will also be corrections to the kink mass from the 1-loop correction to the vacuum energy, which can also be calculated in standard perturbation theory as in \cite{hui}.  These higher loop corrections will be reported in future work.

Our long term goal is to find the monopole operator responsible for confinement in Yang-Mills theory and to show that it is tachyonic as suggested by the paradigm of \cite{thooftconf,mandconf}.  To arrive there, we will try to extend our construction to $\mathcal{N}$=2 Super Yang-Mills where there is a continuous deformation from the semiclassical to the condensing monopole \cite{sw2}.  Here we will start by reproducing the one-loop corrections already found in Ref.~\cite{wimmer}.

\section* {Acknowledgement}

\noindent
JE is supported by the CAS Key Research Program of Frontier Sciences grant QYZDY-SSW-SLH006 and the NSFC MianShang grants 11875296 and 11675223.   JE also thanks the Recruitment Program of High-end Foreign Experts for support.

%%%%%%%%%%%%%%%%%%%%%%%%%%%%%%%%

\end{document}